\documentclass[12pt]{article}

\usepackage[margin=1in]{geometry}
\usepackage{xcolor}
\usepackage[numbers,sort&compress]{natbib}
\usepackage[
  colorlinks=true,
  citecolor=blue,
  linkcolor=blue,
  urlcolor=blue
]{hyperref}
\usepackage{authblk}

\usepackage[T1]{fontenc}
\usepackage{graphicx}
\usepackage{amsmath}
\usepackage{amssymb}
\usepackage{lmodern}
\usepackage{subcaption}
\usepackage{microtype}

\newcommand{\maybegraphic}[2][0.78\linewidth]{%
    \IfFileExists{#2}{%
        \includegraphics[width=#1]{#2}%
    }{%
        \fbox{\parbox{0.72\linewidth}{\centering
        Figure file \texttt{\detokenize{#2}} has not yet been generated.\par
        Run \texttt{appendix\_benchmark\_plots.py} to create it.}}%
    }%
}

\newcommand{\Nb}{N_{\mathrm b}}
\newcommand{\Fdense}{F_{\mathrm{dense}}}
\newcommand{\Flr}{F_{\mathrm{LR}}}
\newcommand{\Fttm}{F_{\mathrm{TTM}}^{(\mathrm{std})}}
\newcommand{\Fph}{F_{\mathrm{ph}}}

\title{Parameter-efficient neural computing with programmable photonic meshes}

\author[1]{Artem Popov}
\author[2,1]{Stanislav Straupe}
\author[2,1]{Mikhail Saygin\thanks{Corresponding author: \href{mailto:saygin@physics.msu.ru}{saygin@physics.msu.ru}}}

\affil[1]{Faculty of Physics, M. V. Lomonosov Moscow State University, Leninskie Gory 1, Moscow 119991, Russia}
\affil[2]{Sber Quantum Technologies Center, Kutuzovski Prospect 32, Moscow 121170, Russia}

\date{}

\begin{document}
\maketitle

\begin{abstract}
Memory capacity and bandwidth are becoming major constraints in neural-network computing, motivating parameter-efficient representations that reduce model storage and data movement. Tensor-network methods already enable substantial compression of neural networks, but their structured operations are often poorly matched to conventional processors optimized for dense linear algebra. Such representations can be implemented in photonic hardware, although this typically requires optical architectures tailored to a particular factorization. Here we propose a different approach based on programmable photonic meshes. Compact electronic parameters determine the interferometer phases, while optical interference produces a dense, generally high-rank transformation. Since the parameterization remains electronic, its dimensionality and structure can be changed without modifying the optical circuit. Numerical experiments in end-to-end training demonstrate that the proposed photonic layers match or outperform strongly compressed tensor-train and brickwork layers, often with fewer trainable parameters. Experiments in parameter-efficient fine-tuning of a language model also show advantages over established digital adapters. The reduced trainable state lowers persistent memory requirements and offers the potential to reduce data movement and bandwidth demands.
\end{abstract}


\section{Introduction}

Modern neural networks rely heavily on large linear transformations, whose storage and execution account for a substantial fraction of memory traffic and computational cost. Growing model sizes have made memory capacity and bandwidth increasingly important constraints, motivating extensive work on parameter-efficient representations, including low-rank factorizations, tensor networks, structured matrices, and parameter-efficient fine-tuning methods~\cite{novikov2015tensorizing,tn_review,tarasov2026_tn_review,hu2021lora,BOFT}. Such constructions can substantially reduce the number of trainable parameters and thereby lower the memory required for weights, gradients, and optimizer states.

Parameter efficiency, however, does not necessarily translate into efficient execution on conventional digital processors. Structured layers commonly replace large dense matrix multiplications with sequences of smaller matrix products and tensor contractions. Depending on the representation and implementation, these operations may suffer from limited hardware utilization, additional data rearrangements, and intermediate memory traffic~\cite{dao2022monarch,lv2022realistic,deng2019tie,gong2023ette}. Carefully designed contraction algorithms and specialized hardware can mitigate these limitations~\cite{qiu2024computebetterspent}. Nevertheless, the compactness of a representation and the efficiency of its execution remain distinct considerations.

Programmable photonics offers another way to implement linear transformations. Optical propagation and interference perform matrix multiplication directly, without decomposing the operation into electronic multiply--accumulate instructions. Multimode interferometers composed of programmable two-mode elements can realize arbitrary unitary transformations and have become established building blocks for photonic computing~\cite{reck1994unitary,clements2016interferometers,carolan2015universal,harris2018pnp,shen2017nanophotonic}. Recent numerical studies have also investigated diffractive optical multiplication within transformer self-attention, reporting tolerance to optical errors and improved task performance with additional trainable optical elements~\cite{SamaraPhotonics}.

Several approaches have combined parameter-efficient neural representations with photonic hardware. Xiao \textit{et al.} proposed tensorized optical neural networks in which tensor-train decompositions are implemented using cascaded photonic cores~\cite{xiao2021tonn}. Other approaches reduce the number of physical optical components by restricting the class of implemented transformations. These include compact interferometric meshes~\cite{tian2022compact}, butterfly-style optical architectures~\cite{feng2022osnn}, and photonic tensor cores based on structured matrix compression~\cite{ning2025cirptc}. Such architectures can substantially reduce optical hardware requirements, but their physical organization is closely related to the chosen matrix structure or factorization.

Here we pursue a different approach, retaining a general programmable multimode interferometer while introducing parameter efficiency in its electronic control representation. Rather than optimizing all interferometer phases independently, we generate them from a much smaller set of trainable parameters. Optical interference then produces a dense, generally high-rank transformation despite the compact electronic representation. Since the parameterization remains electronic, its dimensionality and structure can be changed without modifying the optical circuit. The proposed approach thus reduces the independent trainable model state while retaining the full set of physical phase controls.

We evaluate this approach in both training from scratch and parameter-efficient fine-tuning. Numerical experiments show that the proposed photonic layers can match or outperform strongly compressed tensor-train and brickwork layers over substantial ranges of parameter budgets. Experiments on language-model adaptation also demonstrate favorable performance compared with established digital adapters, including LoRA and BOFT, at smaller trainable parameter counts. These results demonstrate that programmable optical interference can generate compactly parameterized neural transformations without requiring an explicit tensor decomposition of the realized operator.

\section{Parameter-efficient photonic matrix multiplication}
\label{sec:photonic_layer}

We consider photonic implementations of linear neural-network layers, focusing on the relation between their trainable parameters and the optical transformations they realize. The distinction between two conventional photonic architectures motivates our approach: elementwise multipliers, in which individual controls determine individual matrix entries, and interferometric meshes, in which the transfer matrix depends collectively on all phase settings.

\subsection{Universal photonic multipliers}
\label{sec:universal_mvm}

Photonic processors implement linear transformations through optical propagation and interference. For a real-valued linear layer with transfer matrix $W(\Psi)\in\mathbb{R}^{M\times M}$, where $\Psi$ denotes the programmable settings, the transformation of an input vector $\mathbf{x}$ is
    \begin{equation}
        \mathbf{y}=W(\Psi)\mathbf{x}.
        \label{eqn:linear_map}
    \end{equation}
In neural networks, the same transformation is generally applied to multiple activation vectors. Collecting $N_{\mathrm b}$ input vectors as rows of $X\in\mathbb{R}^{N_{\mathrm b}\times M}$, the corresponding batched operation becomes
    \begin{equation}
        Y=XW^{\mathsf T}(\Psi).
        \label{eqn:batched_linear_map}
    \end{equation}
The optical processor performs the individual vector transformations, while $N_{\mathrm b}$ specifies the number of vectors processed with the same configuration.

    \begin{figure}[t]
        \centering
        \includegraphics[width=0.95\linewidth]
        {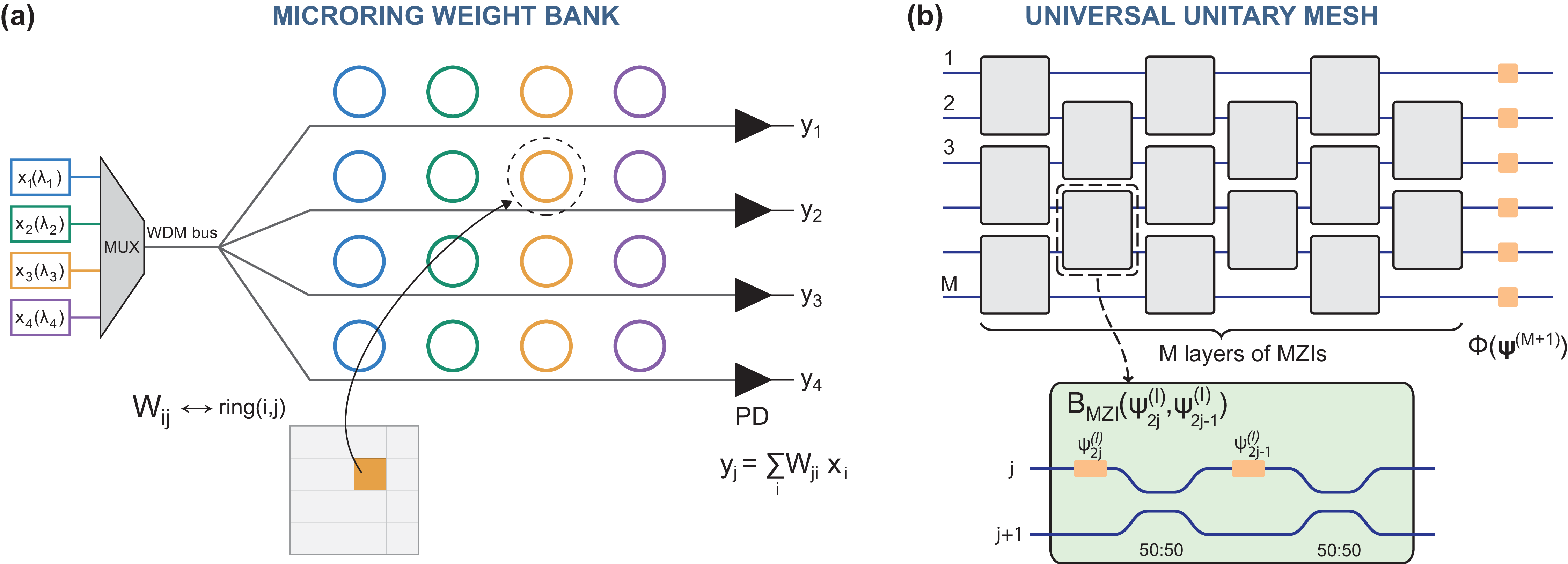}
        \caption{\textbf{Two architectures for programmable photonic matrix multiplication.} (a) Elementwise architecture, illustrated by a microring weight bank. Individual optical controls determine the corresponding matrix entries, making a prescribed dense matrix straightforward to program. (b) Universal interferometric mesh based on the Clements architecture~\cite{clements2016interferometers}. The transfer matrix depends collectively on the phase settings through repeated optical interference, without a direct correspondence between individual phases and matrix entries.}\label{fig:multipliers_types}
    \end{figure}

Elementwise architectures, such as the microring weight bank shown in Fig.~\ref{fig:multipliers_types}(a), provide a direct way to implement a prescribed matrix~\cite{zhang2022microring}. Each matrix element is controlled by a corresponding optical element, schematically as $W_{mn}=f(\Psi_{mn})$. Consequently, once the entries of $W$ are known, they can be programmed without decomposing the matrix into optical transformations.

However, such direct programmability offers no intrinsic advantage for constructing a dense transformation from a compact set of trainable parameters. The compact representation must first be evaluated electronically to obtain all $M^2$ matrix entries, which are then transferred to the optical controls. Thus, the optical hardware performs the multiplication, while the construction of the dense matrix remains digital.

Universal interferometric meshes provide a different possibility. The rectangular Clements architecture, shown in Fig.~\ref{fig:multipliers_types}(b), consists of tunable Mach--Zehnder interferometers and phase shifters~\cite{clements2016interferometers}. For $M$ optical modes, the complete mesh contains $M^2$ independent phase controls and can realize an arbitrary unitary transformation $U_{\mathrm{MZI}}(\Psi)$.

Unlike elementwise multipliers, the mesh has no direct correspondence between individual phase settings and matrix entries. Its transfer matrix depends in a complicated nonlinear way on the phases through repeated splitting, phase accumulation, and interference. This makes the programming of a prescribed unitary matrix computationally more involved, since its decomposition into interferometer settings must first be determined~\cite{nemkov_complexity}. For parameter-efficient neural layers, however, the same property becomes useful. Instead of calculating the phases required for a prescribed dense matrix, we generate them directly from a compact set of trainable parameters. The interferometer then produces the resulting dense transformation through optical interference. This principle forms the basis of the construction introduced below.

\subsection{Parameter-efficient photonic multiplier}

    \begin{figure}[t]
        \centering
    \includegraphics[width=0.95\linewidth]
        {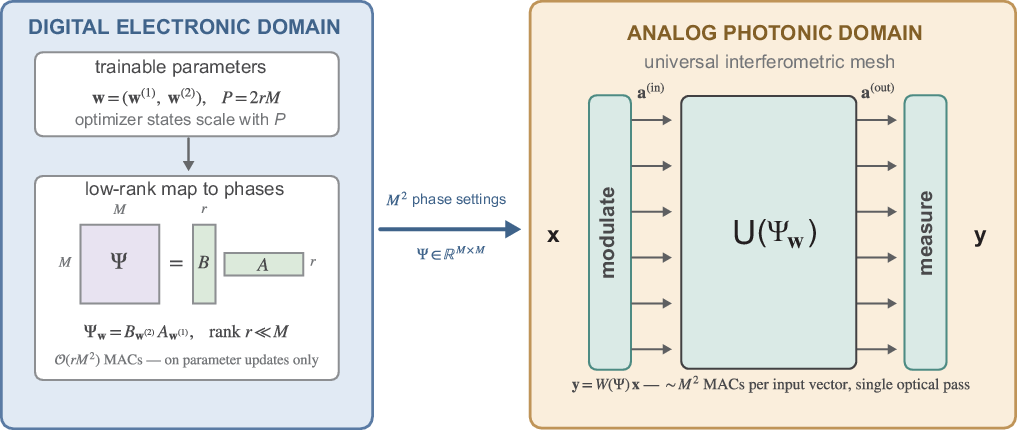}
        \caption{\textbf{Parameter-efficient photonic matrix multiplier.} The electronic parameters $\boldsymbol{w}$ define a low-rank latent matrix $Z_{\boldsymbol{w}}=BA$, from which the $M^2$ interferometer phases are generated through the nonlinear map $\Psi_{\boldsymbol{w}}=2\pi\sigma(Z_{\boldsymbol{w}})$. The programmed mesh implements a dense optical transformation, while the number of trainable parameters is limited to $P=2Mr$.} \label{fig:pe_approach}
    \end{figure}

Figure~\ref{fig:pe_approach} illustrates the proposed hybrid layer. The universal interferometer retains its $P_{\mathrm{mesh}}=M^2$ physical phase controls, but these controls are not treated as independent trainable variables. Instead, their values are generated from a compact electronic parameter vector $\boldsymbol{w}$ with $P\ll P_{\mathrm{mesh}}$ components. The persistent trainable state, including the corresponding gradients and optimizer variables, therefore scales with $P$ rather than $M^2$.

For the implementation considered here, the electronic parameters are arranged into two matrices, $A\in\mathbb{R}^{r\times M}$ and $B\in\mathbb{R}^{M\times r}$, which define the latent control matrix
    \begin{equation}
        Z_{\boldsymbol{w}}=BA,
        \qquad
        \operatorname{rank}(Z_{\boldsymbol{w}}) = r.
        \label{eqn:latent_control}
    \end{equation}
The interferometer phases are generated by the elementwise nonlinear mapping
    \begin{equation}
        \Psi_{\boldsymbol{w}}
        =
        2\pi\,\sigma(Z_{\boldsymbol{w}}),
        \qquad
        \sigma(t)=\frac{1}{1+e^{-t}},
        \label{eqn:phase_map}
    \end{equation}
where the $M^2$ phase values are arranged into an $M\times M$ array for convenience.

Importantly, the rank constraint applies to the latent matrix $Z_{\boldsymbol{w}}$, not to the physical phase matrix $\Psi_{\boldsymbol{w}}$. The nonlinear map generally does not preserve rank, and the resulting optical transformation $U_{\mathrm{MZI}}(\Psi_{\boldsymbol{w}})$ is dense and high rank. The compact electronic representation therefore does not impose a corresponding low-rank constraint on the neural transformation, although it restricts the family of accessible operators.

The number of independent trainable parameters is
    \begin{equation}
        P=2Mr,
        \label{eqn:photonic_params}
    \end{equation}
with the rank $r$ controlling the parameter budget without changing the physical interferometer. Constructing a new configuration requires the electronic product $BA$, which costs approximately $2rM^2=MP$ FLOPs. This cost is incurred once per reprogramming and can be amortized over multiple input vectors. Its comparison with conventional dense, low-rank, and tensor-train execution is given in Appendix~\ref{app:electronic_complexity}. The comparison excludes optical propagation and the physical cost of loading the phase settings.

The interferometer acts on complex optical amplitudes, whereas the neural-network layers considered here use real-valued vectors. We employ two representations. In encoding (i), each optical mode carries one real vector component, and measuring a single output quadrature gives
    \begin{equation}
        W(\Psi)=\operatorname{Re}U_{\mathrm{MZI}}(\Psi).
        \label{eqn:encoding_1}
    \end{equation}
In encoding (ii), both quadratures carry independent real components, so an $M$-mode interferometer implements an orthogonal transformation on a $2M$-dimensional real vector. The first representation is used for the MNIST and CIFAR-100 benchmarks, while the second is used for the photonic adapter in the Qwen3 experiment. Details of the Clements mesh and the optical encoding and detection schemes are given in Appendices~\ref{app:mesh_details} and~\ref{app:optical_encoding}.

\section{Baseline parameter-efficient layers}\label{sec:baseline_layers}

\begin{figure}
        \centering
        \includegraphics[width=0.95\linewidth]{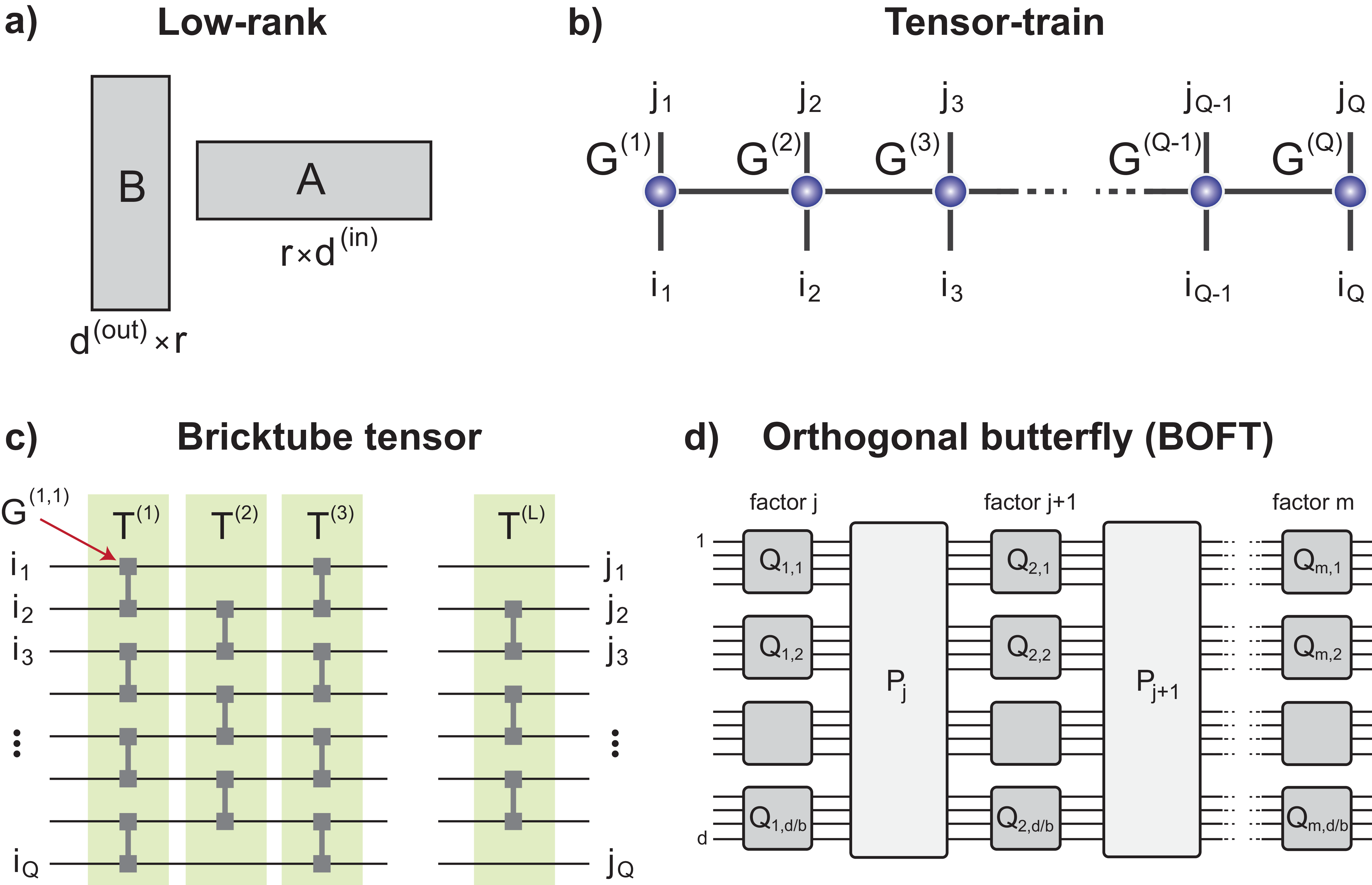}
        \caption{\textbf{Digital parameter-efficient layers used as baselines}:
        (a) low-rank layer, (b) tensor-train layer, (c) brickwork tensor-network
        layer, and (d) orthogonal butterfly factor. In end-to-end training, the
        layers of panels (a)--(c) replace the dense linear layers of the network;
        in fine-tuning, the low-rank construction (a) serves as the additive
        adapter (LoRA) and the butterfly factor (d) as the multiplicative adapter
        (BOFT).}
        \label{fig:digital_layers}
    \end{figure}

We compare the proposed photonic layer with four parameter-efficient baselines, summarized in Fig.~\ref{fig:digital_layers}. For end-to-end training, low-rank (LR)~\cite{hu2021lora}, tensor-train (TT)~\cite{oseledets2011tt}, and brickwork (BW)~\cite{rieser_brickwork} layers replace the corresponding dense linear transformations. For fine-tuning, we compare the proposed multiplicative photonic adapter with the additive low-rank adapter LoRA~\cite{hu2021lora} and the multiplicative orthogonal adapter BOFT~\cite{BOFT}. Detailed definitions and parameter counts of all baseline constructions are given in Appendix~\ref{app:baseline_layers}.

The LR baseline factorizes a square weight matrix as
    \begin{equation}
        W^{(\mathrm{LR})}=BA,
        \qquad
        B\in\mathbb{R}^{M\times r},
        \quad
        A\in\mathbb{R}^{r\times M},
        \label{eqn:low_rank}
    \end{equation}
using $P_{\mathrm{LR}}=2rM$ trainable parameters. The same two-factor construction is used in the proposed photonic layer to generate the \emph{latent control matrix}, rather than to factorize the realized neural transformation. Comparison with the conventional LR layer therefore isolates the effect of the interferometric parameter-to-weight map from that of the underlying compact electronic representation. In fine-tuning, the low-rank construction is applied additively,
    \begin{equation}
        W_{\mathrm{adapt}}
        =
        W_0+\Delta W^{(\mathrm{LR})},
        \label{eqn:additive_adapter}
    \end{equation}
which corresponds to LoRA.

The TT~\cite{oseledets2011tt} and BW~\cite{rieser_brickwork} baselines instead represent the weight matrix by structured tensor networks. TT uses a chain of low-rank tensor cores, whereas BW uses local four-way cores arranged in alternating nearest-neighbor layers. Their parameter budgets are varied through the TT ranks and the brickwork depth, respectively. Both constructions provide compact structured alternatives to a dense matrix without imposing the simple rank constraint of Eq.~\eqref{eqn:low_rank}.

For multiplicative fine-tuning we use BOFT as implemented in the Hugging Face PEFT library. BOFT represents the adapter by block-orthogonal factors connected through butterfly permutations. In our experiments, its parameter budget is varied through the BOFT block size while the remaining BOFT configuration is held fixed. The reported parameter counts are the actual trainable parameter counts of the instantiated PEFT adapters. With the convention used here,
    \begin{equation}
        W_{\mathrm{adapt}}=W_0 R,
        \qquad
        R^{\mathsf T}R=I ,
        \label{eqn:boft_mult}
    \end{equation}
so BOFT provides a structured, orthogonal, multiplicative baseline for the proposed photonic adapter. Unlike LoRA, its parameter efficiency is imposed by architectural structure rather than by a rank constraint.

    \begin{figure}
        \centering
        \includegraphics[width=0.95\linewidth]{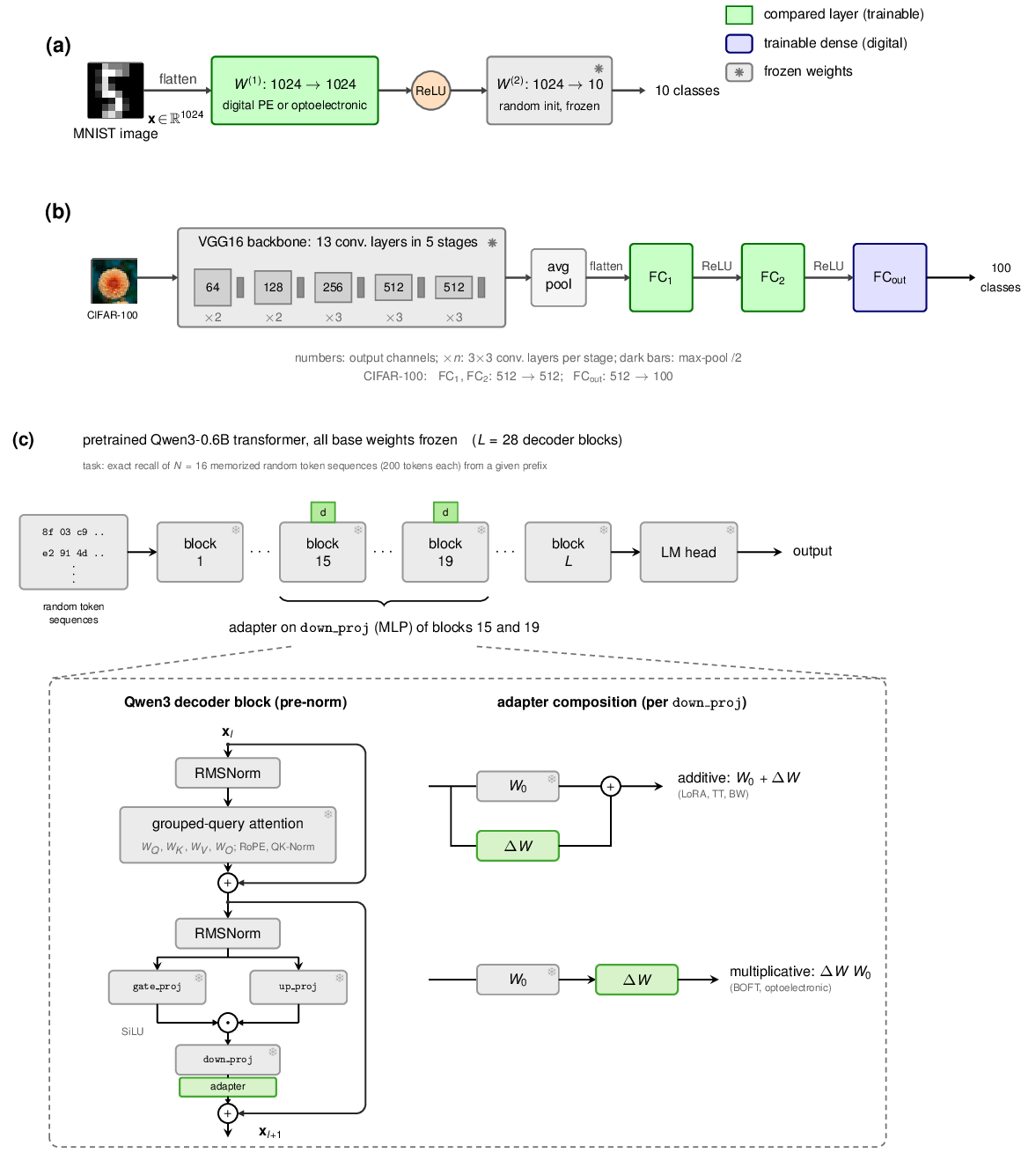}
        \caption{\textbf{Neural-network architectures used in the benchmarks.} (a) Two-layer MNIST-10 classifier with a trainable $1024\times1024$ layer followed by ReLU and a frozen, randomly initialized output layer. (b) CIFAR-100 classifier with a frozen VGG16 backbone and a trainable classification head. The compared parameterizations are applied to two $512\times512$ layers, while the final classification layer remains digital matrix with independent entries. (c) Qwen3-0.6B transformer with parameter-efficient adapters in the \texttt{down\_proj} modules of decoder blocks 15 and 19. The pretrained weights remain frozen. The inset illustrates additive and multiplicative adaptation of the pretrained transformations.}\label{fig:architectures}
    \end{figure}

\section{Results}\label{sec:results}

We evaluate the proposed parameter-efficient photonic layer in three neural-network benchmarks, covering both training from random initialization and adaptation of pretrained models (Figure~\ref{fig:architectures}). The first benchmark considers a two-layer MNIST-10 classifier, where the compared parameterizations replace a single dense linear transformation. The second benchmark examines classification on CIFAR-100 using a frozen pretrained VGG16 backbone and a trainable classification head containing two parameter-efficient layers. In both cases, we compare the photonic implementation with conventional LR, TT, and BW layers, using the number of trainable parameters as the principal measure of model complexity. The third benchmark considers parameter-efficient fine-tuning of Qwen3-0.6B through a controlled sequence-memorization task, comparing the proposed photonic adapter with LoRA, TT, and structured orthogonal adaptation methods.

The numerical experiments also examine the sensitivity of the photonic parameterization to finite phase resolution. Alongside the standard FP32 implementation, we consider 8-bit quantization, motivated by experimental progress in high-precision and quantized photonic computing~\cite{zhang2022microring,han2026quantized}. For the photonic layer, the latent matrix $Z=BA$ and the subsequent sigmoid transformation are evaluated in FP32, while quantization is applied to the resulting interferometer phase settings. During training, gradients through the quantization operation are estimated using the straight-through estimator (STE)~\cite{jacob2018quantization}.

For the MNIST-10 and CIFAR-100 benchmarks, each configuration is trained with ten random seeds. The reported classification errors are the means over these runs, with error bars indicating one standard deviation. The sequence-memorization benchmark instead uses a single fixed set of random token sequences, without repeated runs. Consequently, its results characterize the parameter efficiency of the compared adapters for the investigated task instance and training settings. All numerical experiments for training and fine-tuning neural networks are conducted in PyTorch and executed on an NVIDIA A100 GPU with $40$ GB of memory.

\subsection{End-to-end training: two-layer MNIST classifier}\label{sec:results_mnist}

The first benchmark examines the performance of parameter-efficient layers trained from random initialization. Each MNIST image is represented by a $1024$-dimensional input vector and processed by a two-layer classifier, as illustrated in Figure~\ref{fig:architectures}(a). The first $1024\times1024$ transformation is implemented using the proposed photonic layer or one of the digital baselines, followed by a ReLU activation and a frozen, randomly initialized output layer. The resulting classification error therefore provides a direct comparison of the representations used for the trainable transformation.

\begin{figure}
        \centering
        \includegraphics[width=0.75\linewidth]{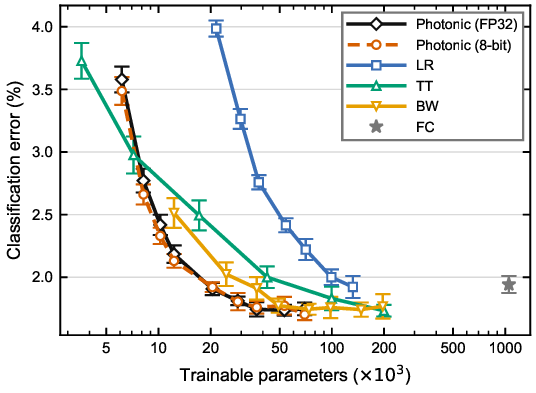}
        \caption{\textbf{End-to-end training on MNIST-10.} Mean test classification error as a function of the number of trainable parameters for the photonic, low-rank (LR), tensor-train (TT) and brickwork (BW) layers, compared with the dense fully connected (FC) reference. Full-precision and $8$-bit results are shown for the photonic layer. Error bars indicate one standard deviation over ten random seeds.} \label{fig:mnist}
    \end{figure}

Figure~\ref{fig:mnist} shows the mean test error and standard deviation over ten random seeds as functions of the number of trainable parameters. The dense reference reaches $(1.94\pm0.07)\%$ error with approximately $1.05\times10^6$ parameters. At the smallest parameter budgets, the TT layer provides a more accurate representation than the photonic layer. However, the curves cross at approximately $P=8\times10^3$ parameters, beyond which the photonic layer exhibits lower mean classification errors throughout the overlapping parameter range. With $2.05\times10^4$ parameters, the full-precision photonic layer already matches the dense reference, reaching $(1.91\pm0.05)\%$ error. Increasing the parameter budget to $3.69\times10^4$ reduces the error to $(1.74\pm0.06)\%$, while the lowest mean error of $(1.74\pm0.04)\%$ is obtained with $5.32\times10^4$ parameters.

The tensor-network baselines approach similar accuracies at larger parameter budgets. The brickwork (BW) layer reaches $(1.77\pm0.06)\%$ error with $4.92\times10^4$ parameters and $(1.74\pm0.04)\%$ with $7.37\times10^4$ parameters. The TT layer continues to improve with increasing rank, eventually reaching $(1.73\pm0.05)\%$ at approximately $2.00\times10^5$ parameters. Thus, both tensor-network representations attain accuracy comparable to the photonic layer, but the latter reaches this performance range with approximately two times fewer parameters than BW and 5.4 times fewer than TT. The differences between the lowest errors are small compared with their standard deviations. The conventional low-rank layer is less efficient in this benchmark, reaching its lowest reported error of $(1.92\pm0.09)\%$ with $1.32\times10^5$ parameters.

Finite phase resolution has little effect on the observed performance. In the $8$-bit photonic implementation, the latent matrix $Z=BA$ and the sigmoid transformation are evaluated in FP32, while only the resulting interferometer phases are quantized. At $3.69\times10^4$ parameters, the quantized layer reaches $(1.76\pm0.05)\%$ error, compared with $(1.74\pm0.06)\%$ for full precision. The lowest mean error of the 8-bit implementation is $(1.7\pm0.05)\%$ at $6.96\times10^4$ parameters. Thus, within the statistical variability of the numerical experiments, phase quantization to 8 bits produces no systematic degradation in classification accuracy.

\subsection{End-to-end training on CIFAR-100}
\label{sec:results_vgg}

    \begin{figure}
        \centering
        \includegraphics[width=0.75\linewidth]{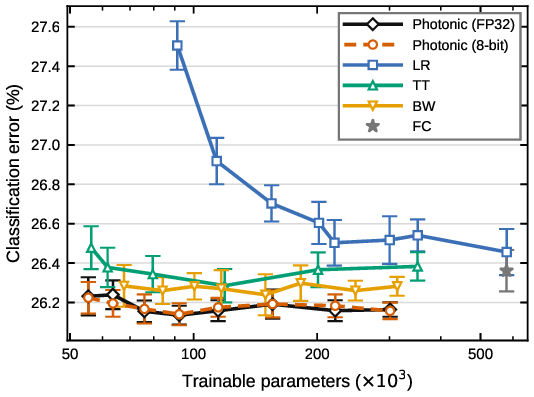}
        \caption{\textbf{End-to-end training on CIFAR-100.} Mean test classification error as a function of the total number of trainable parameters in the classification head for the photonic, low-rank (LR), tensor-train (TT) and brickwork (BW) layers, together with the dense fully connected (FC) reference. Full-precision and $8$-bit results are shown for the photonic layer. Error bars indicate one standard deviation over ten random seeds.}\label{fig:cifar}
    \end{figure}

The second benchmark considers classification on CIFAR-100 using a pretrained VGG16 network with frozen convolutional layers, as illustrated in Figure~\ref{fig:architectures}(b). The resulting $512$-dimensional feature vectors are processed by a trainable classification head containing two $512\times512$ transformations, each followed by LayerNorm and ReLU, and a final dense $512\times100$ output layer. The first two transformations are replaced by the parameter-efficient constructions under comparison. Consequently, the experiment tests their performance within a multilayer neural network while retaining the same pretrained feature representation.

Figure~\ref{fig:cifar} shows the test classification error averaged over ten random seeds. The dense classification head reaches $(26.36\pm0.11)\%$ error with $5.79\times10^5$ trainable parameters. The proposed photonic head already achieves $(26.23\pm0.1)\%$ with $5.54\times10^4$ parameters, corresponding to approximately ten times fewer trainable parameters for the complete classification head. Increasing the photonic rank produces a modest further improvement, with the lowest mean error of $(26.14\pm0.05)\%$ obtained at $9.23\times10^4$ parameters.

The TT baselines exhibit different dependencies on the parameter budget. The brickwork (BW) layer already reaches $(26.28\pm0.11)\%$ error with $6.77\times10^4$ parameters, and increasing its depth produces only modest changes in accuracy. Its lowest mean error of $(26.24\pm0.10)\%$ is obtained with $1.50\times10^5$ parameters. The tensor-train layer, in contrast, improves with increasing rank and reaches its lowest error of $(26.29\pm0.09)\%$ at $1.19\times10^5$ parameters. Thus, both tensor-network representations achieve competitive classification performance, although the photonic layer reaches comparable accuracy with considerably fewer parameters. In particular, the photonic head already achieves $(26.23\pm0.10)\%$ error with $5.54\times10^4$ parameters, approximately 2.7 times fewer than required for the best BW result. The differences between the lowest classification errors remain small relative to the variability across random seeds and should not be overinterpreted. The conventional low-rank layer is considerably less efficient, requiring substantially larger parameter budgets while remaining less accurate than the photonic and tensor-network layers.

The results are also remarkably insensitive to phase quantization. The full-precision and 8-bit photonic curves remain close throughout the investigated parameter range. At $9.23\times10^4$ parameters, the corresponding errors are $(26.14\pm0.05)\%$ and $(26.14\pm0.06)\%$, respectively. Therefore, the numerical results indicate that the proposed photonic parameterization retains its classification performance when the programmed phase settings are restricted to 8-bit precision.

\subsection{Fine-tuning: sequence memorization with Qwen3}\label{sec:results_memorization}

    \begin{figure}
        \centering
        \includegraphics[width=0.95\linewidth]{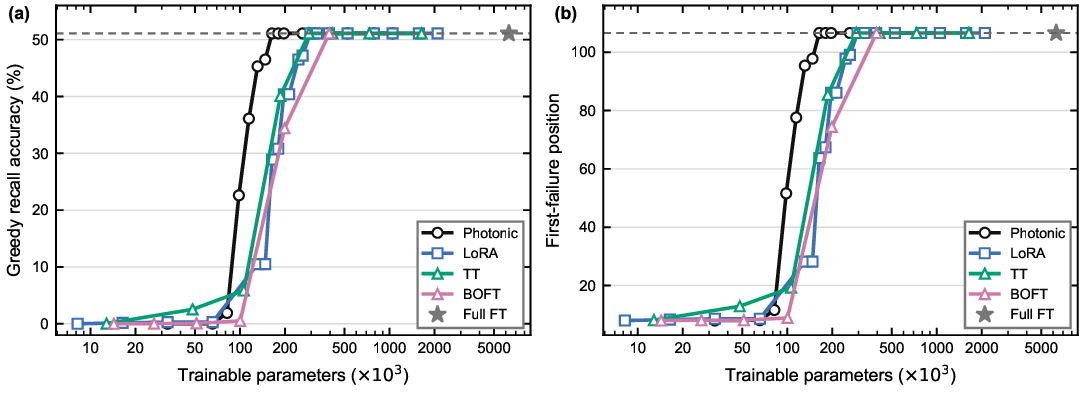}
        \caption{\textbf{Parameter-efficient sequence memorization  with Qwen3-0.6B.}
        Memorization performance as a function of the number of trainable parameters for the photonic adapter, LoRA, tensor-train (TT), BOFT, and OFT. (a) Greedy recall accuracy. (b) Mean first-failure position. Horizontal dashed lines indicate the performance of full fine-tuning (Full FT). Each point corresponds to one numerical experiment on the same fixed set of random sequences.}
        \label{fig:memo}
    \end{figure}

The final benchmark evaluates parameter-efficient fine-tuning of the pretrained Qwen3-0.6B language model~\cite{qwen} through a controlled sequence-memorization task~\cite{lora_memory}. The model is trained on $N=16$ random token sequences of length $\ell=200$, with trainable adaptations introduced into the \texttt{down\_proj} modules of transformer blocks 15 and 19. All remaining pretrained parameters are kept frozen.  Further details of the sequence-memorization protocol, adapter implementations, and evaluation metrics are provided in Appendix~\ref{app:qwen_memorization}.

The comparison includes the proposed photonic adapter, LoRA, TT adaptation, BOFT, and full fine-tuning of the two selected projection matrices. LoRA and TT introduce additive updates to the rectangular pretrained weight matrices. The TT adapter uses a four-core representation with variable internal rank. In contrast, the photonic and orthogonal adapters use multiplicative transformations. The photonic implementation employs $M=512$ optical modes and the quadrature encoding to act on $1024$-dimensional real representations.

Figure~\ref{fig:memo} shows greedy recall accuracy and the mean first-failure position as functions of the total number of trainable parameters. All methods exhibit a pronounced transition from negligible recall to substantially improved memorization as the parameter budget increases. For the fixed sequence set considered here, the highest observed greedy recall accuracy is approximately $51.1\%$, with a mean first-failure position of $106.6$. This level is also reached by full fine-tuning of the two selected dense matrices.

The principal difference between the methods is the parameter budget required to reach this performance. The photonic adapter reaches the observed plateau at $P\approx1.64\times10^5$ trainable parameters. The corresponding counts are approximately $2.89\times10^5$ for TT, $2.95\times10^5$ for LoRA, and $3.95\times10^5$ for BOFT. The full fine-tuning reference requires $6.29\times10^6$ parameters. Consequently, the photonic adapter reaches the same observed memorization level with approximately $1.8$ times fewer trainable parameters than TT and LoRA, $2.4$ times fewer than BOFT, and $38$ times fewer than the dense fine-tuning reference.

The advantage is also apparent below saturation. With approximately $1.31\times10^5$ parameters, the photonic adapter reaches $45.3\%$ greedy recall accuracy and a mean first-failure position of $95.4$. At essentially the same parameter count, LoRA reaches $10.5\%$ recall and a mean first-failure position of $28.2$. The TT adapter approaches comparable recall only at larger parameter budgets. Thus, the photonic parameterization not only reaches the observed plateau with fewer trainable variables but also exhibits a transition toward successful memorization at smaller parameter counts.

These results provide evidence that the compact electronic parameterization of a programmable interferometer can serve as an effective alternative to conventional low-rank, tensor-network, and structured orthogonal adapters. However, the memorization benchmark uses a single fixed set of random sequences without repeated training runs. The reported thresholds should therefore be interpreted as results for the investigated task instance and optimization settings, rather than as universal estimates of the memorization capacity of the respective adapter families.

\section{Discussion and Outlook}

The numerical results demonstrate that programmable photonic interferometers can provide parameter-efficient neural transformations competitive with strongly compressed digital representations. In end-to-end training, the proposed photonic layer matches the dense reference on MNIST-10 with approximately $50$ times fewer trainable parameters, while on CIFAR-100 the corresponding reduction for the classification head is about tenfold. TT and BW layers provide strong digital baselines and eventually reach comparable classification accuracy. However, the photonic layer generally achieves this performance with fewer trainable parameters. In fine-tuning of Qwen3-0.6B, the photonic adapter reaches the observed memorization performance of the reference obtained by optimizing the two selected dense projection matrices with approximately 1.8 times fewer parameters than TT and LoRA, and 2.4 times fewer than BOFT. The latter results concern a single fixed set of random sequences and therefore do not establish general limits on the memorization capacity of these representations.

The distinction from conventional structured layers lies in how the dense transformation is generated. Tensor-network representations express a large operator through contractions of smaller tensors~\cite{tn_review,tarasov2026_tn_review}. In the proposed approach, a compact electronic representation determines the phase settings of a multimode interferometer, while the resulting transformation is generated through optical interference. The low-rank constraint applies to the latent electronic matrix $Z=BA$, rather than to the phase settings or the optical transfer matrix. Consequently, the realized neural transformation is not restricted to rank $r$, although the compact parameterization limits the family of accessible transformations. Once the phase settings are established, the interferometer implements the resulting transformation without repeatedly evaluating the compact electronic representation. The electronic cost of constructing a new configuration and its amortization over multiple input vectors are analyzed in Appendix~\ref{app:electronic_complexity}.

Reducing the number of trainable parameters directly lowers the persistent electronic memory required for weights, gradients, and optimizer states. This benefit is particularly relevant as memory capacity and data movement become increasingly important constraints in neural-network computing~\cite{gholami2024memorywall,sebastian2020memory}. The numerical results also show that restricting the interferometer phase settings to 8-bit precision does not produce a systematic degradation in classification accuracy. However, this observation concerns phase quantization applied after evaluating the electronic parameterization in FP32 and does not account for optical loss, phase noise, or calibration errors. Practical hardware advantages therefore remain to be established experimentally. The universal interferometer retains its $M^2$ physical phase controls, and constructing their settings may require temporary memory scaling as $M^2$. Furthermore, the computational estimates in Appendix~\ref{app:electronic_complexity} exclude phase loading, device settling, calibration, and the optical input and output interfaces. The achievable reductions in latency, energy consumption, and memory bandwidth will depend on these additional costs and on the number of input vectors processed between successive updates of the optical configuration.

The present approach is complementary to photonic architectures that obtain parameter efficiency by restricting the optical transformation itself. Xiao \textit{et al.} proposed tensor-train decompositions implemented using cascaded photonic cores~\cite{xiao2021tonn}, while other approaches employ compact interferometric networks, butterfly architectures, or structured matrix compression~\cite{tian2022compact,feng2022osnn,ning2025cirptc}. Such designs can reduce the number of optical components and the associated hardware complexity, but their physical organization is closely related to the chosen factorization or matrix structure. Here parameter efficiency is introduced in the electronic representation, while the interferometer retains its full set of programmable elements. Thus, different parameter budgets and functional forms of the electronic map can be explored on the same physical processor. Combining compact electronic representations with optical architectures of reduced complexity offers another direction for future work.

Scaling the optical processor remains a central challenge. Neural-network layers commonly operate on vectors with hundreds or thousands of components, whereas experimentally demonstrated universal interferometers remain considerably smaller~\cite{taballione2023universal,hua2025photonic,Song_64x64_mzi}. Tiling provides one route to larger effective transformations by partitioning the matrix into smaller blocks that are evaluated sequentially or in parallel and subsequently accumulated~\cite{giamougiannis2023tiled}. Larger mode spaces may also be obtained by exploiting additional optical degrees of freedom. Programmable frequency transformations and processing that combines spatial and frequency modes have already been demonstrated~\cite{zhao2022frequency,honari2024hyperspectral}. Time-bin encoding provides another possibility. Basani \textit{et al.} recently proposed a universal photonic processor in the synthetic time dimension, using dynamically reconfigurable connections to reduce the number of physical components~\cite{BasaniSynthetic}. These approaches offer routes toward larger effective transformations without a proportional increase in spatial ports, although their practical efficiency will depend on optical losses, control overhead, and the resources required for temporal or spectral multiplexing.

The low-rank electronic map considered here represents only one possible parameterization. More expressive representations may improve performance at comparable parameter budgets, while reduced-depth and error-tolerant interferometers may relax the requirements on device size and stability~\cite{ErrorTolerant,RobustPRL,TwoUnitaryPRR,TangLowDepth}. Combining these approaches with realistic models of optical loss, phase errors, and calibration uncertainty represents an important step toward experimental validation.

More broadly, the results demonstrate that programmable optical interference can generate compactly parameterized neural transformations without explicitly factorizing the realized operator into a tensor network. This provides an alternative physical mechanism for achieving parameter efficiency while retaining the flexibility of a general programmable optical processor. Establishing whether these properties translate into practical hardware advantages will require experimental assessment of programming overhead, optical losses, and total energy and latency.

\clearpage
\appendix

\section{Electronic complexity of parameter-efficient linear layers}
\label{app:electronic_complexity}

This Appendix compares the electronic cost of the parameter-efficient
representations considered in the main text.  The comparison is intentionally
restricted to operations that are executed electronically.  For the photonic
layer we count only the matrix product used to construct the latent phase-control
matrix,
\begin{equation}
    Z = BA,
    \qquad
    B\in\mathbb{R}^{M\times r_{\mathrm{ph}}},
    \qquad
    A\in\mathbb{R}^{r_{\mathrm{ph}}\times M},
    \label{eq:app_ph_ba}
\end{equation}
and do not include evaluation of a phase nonlinearity, transfer of the resulting
values to the optical circuit, phase-shifter settling, calibration, or optical
matrix--vector multiplication.  The purpose of the analysis is therefore not
to estimate the end-to-end latency of a photonic accelerator, but to quantify
the electronic arithmetic and memory overhead required to construct a new
photonic layer whenever its trainable parameters are updated.

Throughout this Appendix, one floating-point multiply and one floating-point
addition are counted as two floating-point operations (FLOPs).  We denote by
$\Nb$ the number of input vectors processed with one fixed layer configuration.
During minibatch training, $\Nb$ is naturally identified with the batch size
(or, more generally, with the number of activation vectors that reuse the same
weights before the next parameter update).  During inference, $\Nb$ is a reuse
factor and need not coincide with the software minibatch size.  We use $\Nb$
rather than $B$ for this quantity in order to avoid confusion with the factor
matrix $B$ in Eq.~\eqref{eq:app_ph_ba}.

\subsection{Leading-order FLOP counts}
\label{app:flops}

\paragraph{Dense layer.}
For a dense matrix $W\in\mathbb{R}^{M\times M}$ and a batch
$X\in\mathbb{R}^{\Nb\times M}$, the product
$Y=XW^{\mathsf T}$ requires, to leading order,
    \begin{equation}
        \Fdense = 2\Nb M^2 .
        \label{eq:app_dense_flops}
    \end{equation}
The dense layer contains $P_{\mathrm{dense}}=M^2$ trainable parameters.

\paragraph{Low-rank layer.}
For
    \begin{equation}
        W_{\mathrm{LR}}=B_{\mathrm{LR}}A_{\mathrm{LR}},
        \qquad
        A_{\mathrm{LR}}\in\mathbb{R}^{r_{\mathrm{LR}}\times M},
        \qquad
        B_{\mathrm{LR}}\in\mathbb{R}^{M\times r_{\mathrm{LR}}},
    \end{equation}
the multiplication is evaluated as
    \begin{equation}
        X
        \longrightarrow
        XA_{\mathrm{LR}}^{\mathsf T}
        \longrightarrow
        (XA_{\mathrm{LR}}^{\mathsf T})B_{\mathrm{LR}}^{\mathsf T}.
    \end{equation}
The parameter count and FLOP count are therefore
    \begin{align}
        P_{\mathrm{LR}} &= 2Mr_{\mathrm{LR}},
        \label{eq:app_lr_params}\\
        \Flr &= 4\Nb Mr_{\mathrm{LR}}
              = 2\Nb P_{\mathrm{LR}} .
        \label{eq:app_lr_flops}
    \end{align}

\paragraph{Tensor-train matrix layer.}
The tensor-train (TT) format was introduced by Oseledets~\cite{oseledets2011tt}.
The neural-network baseline used here follows the TT-matrix
representation of Novikov \textit{et al.}~\cite{novikov2015tensorizing}: for
$M=D^Q$, the weight matrix is represented by paired input/output cores
    \begin{equation}
        G^{(q)}
        \in
        \mathbb{R}^{r_{q-1}\times D_{\mathrm{out}}\times
        D_{\mathrm{in}}\times r_q},
        \qquad
        D_{\mathrm{out}}=D_{\mathrm{in}}=D,
    \end{equation}
with $r_0=r_Q=1$.  This four-way paired-core representation is also commonly
called a \emph{tensor-train matrix} (TTM) in the recent hardware and
tensorized-training literature~\cite{tian2026fpga_tensor_transformer}.  We
retain the shorter term ``TT layer'' in the main text, but use ``TTM'' in this
Appendix when discussing its arithmetic complexity in order to distinguish it
from alternative TT factorizations of a matrix into $2Q$ three-way cores.

The number of trainable parameters is
    \begin{equation}
        P_{\mathrm{TTM}}=D^2\sum_{q=1}^{Q} r_{q-1}r_q .
        \label{eq:app_tt_params}
    \end{equation}
Let
    \begin{equation}
        S_r=\sum_{q=1}^{Q} r_{q-1}r_q .
    \end{equation}
For the standard staged contraction of this TT-matrix with an
explicit dense activation vector, the $q$th core contributes, to
leading order, $MD\,r_{q-1}r_q$ scalar multiplications per input vector.
Consequently, for $\Nb$ dense input vectors,
\begin{equation}
    \Fttm
    \simeq
    2\Nb M D
    \sum_{q=1}^{Q} r_{q-1}r_q
    =
    \frac{2\Nb M}{D}\,P_{\mathrm{TTM}},
    \label{eq:app_tt_flops}
\end{equation}
where the factor of two counts one multiplication and one addition as two
FLOPs.  This expression is consistent with the TT-matrix-by-explicit-vector
complexity derived by Novikov \textit{et al.},
$O(d r^2 m\max\{M,N\})$~\cite{novikov2015tensorizing}.  In the square,
uniform setting of the present work,
$d=Q$, $m=D$, and $M=N=D^Q$, which gives
$O(Qr^2DM)$ operations per vector.

The same scaling can be read directly from the recent complexity analysis of
Tian \textit{et al.}~\cite{tian2026fpga_tensor_transformer}.  For uniform mode
size $n$ and rank $r$, their TTM training complexity is proportional to
$3K n^{d+1}((d-2)r^2+2r)$; the factor of approximately three accounts for the
forward and backward contractions.  Identifying
$K=\Nb$, $n=D$, $d=Q$, and $M=D^Q$, the forward contribution scales as
\begin{equation}
    \Nb M D\big((Q-2)r^2+2r\big),
\end{equation}
scalar multiplications.  Since for uniform internal rank
\begin{equation}
    S_r = 2r+(Q-2)r^2,
\end{equation}
this is the same leading multiplication count as
Eq.~\eqref{eq:app_tt_flops} before the multiplication-to-FLOP conversion.

Equation~\eqref{eq:app_tt_flops} should therefore be interpreted as the cost of
the standard paired-core TT-matrix/TTM contraction with dense
activations, which is the representation benchmarked in this work.  It is
not a representation-independent lower bound for every algorithm described as
``tensor train''.  Contraction ordering and hardware scheduling can materially
change the amount of redundant computation, intermediate storage, and data
movement.  TIE~\cite{deng2019tie} derives a computation-efficient TT inference
scheme designed to attain the theoretical minimum number of multiplications
for its TT-format DNN computation, while ETTE~\cite{gong2023ette} further
co-designs the contraction order and hardware dataflow to reduce stage-wise
computation, intermediate storage, and data movement.  Tian \textit{et al.}
also compare TTM with a $2Q$-core TT representation and introduce a
bidirectional TT contraction with a different complexity scaling
~\cite{tian2026fpga_tensor_transformer}.  These results motivate reporting
measured CPU/GPU runtime in addition to analytical FLOP counts.

Finally, the complexity is different if the activation itself is kept in TT
form.  The original TT literature analyzes TT-operator--TT-vector operations
with rank-dependent costs that are not the same as the explicit-dense-vector
case considered here~\cite{oseledets2011tt}.

\paragraph{Photonic layer: electronic construction only.}
For the photonic layer, the persistent trainable state consists of the two thin
matrices in Eq.~\eqref{eq:app_ph_ba},
\begin{equation}
    P_{\mathrm{ph}}=2Mr_{\mathrm{ph}} .
    \label{eq:app_ph_params}
\end{equation}
A new layer configuration requires the matrix--matrix product $BA$, whose
electronic arithmetic cost is
\begin{equation}
    \Fph
    =
    2M^2r_{\mathrm{ph}}
    =
    M P_{\mathrm{ph}} .
    \label{eq:app_ph_flops}
\end{equation}
This cost is paid once per reprogramming and, within the scope of the present
comparison, is independent of $\Nb$.  Optical execution of the resulting
transformation is deliberately excluded.

\subsection{Comparison at matched trainable-parameter count}
\label{app:matched_P}

To separate parameter efficiency from the details of a particular rank choice,
it is useful to compare the three compact representations at a common
trainable-parameter budget,
\begin{equation}
    P_{\mathrm{LR}}
    =
    P_{\mathrm{TTM}}
    =
    P_{\mathrm{ph}}
    \equiv P .
\end{equation}
Equations~\eqref{eq:app_lr_flops},
\eqref{eq:app_tt_flops}, and~\eqref{eq:app_ph_flops} then reduce to
\begin{align}
    \Flr(P,\Nb) &= 2\Nb P ,
    \label{eq:app_lr_matched}\\
    \Fttm(P,\Nb) &= \frac{2\Nb M}{D}P ,
    \label{eq:app_tt_matched}\\
    \Fph(P) &= MP .
    \label{eq:app_ph_matched}
\end{align}
The dense reference remains
\begin{equation}
    \Fdense(\Nb)=2\Nb M^2 .
\end{equation}

Normalizing by the dense FLOP count gives particularly transparent expressions,
\begin{align}
    \frac{\Flr}{\Fdense}
    &=
    \frac{P}{M^2},
    \label{eq:app_lr_norm}\\
    \frac{\Fttm}{\Fdense}
    &=
    \frac{P}{DM},
    \label{eq:app_tt_norm}\\
    \frac{\Fph}{\Fdense}
    &=
    \frac{P}{2\Nb M}.
    \label{eq:app_ph_norm}
\end{align}
Thus the electronic cost of the low-rank and TTM layers grows linearly with the
number of processed vectors, whereas the $BA$ construction cost of the photonic
layer is paid once and is amortized by reuse.

For the common choice $P=2Mr_{\mathrm{ph}}$, Eq.~\eqref{eq:app_ph_norm}
becomes
\begin{equation}
    \frac{\Fph}{\Fdense}
    =
    \frac{r_{\mathrm{ph}}}{\Nb} .
    \label{eq:app_ph_dense_simple}
\end{equation}
The electronic $BA$ construction therefore requires fewer FLOPs than evaluating
the corresponding dense layer on the same set of input vectors once
\begin{equation}
    \Nb > r_{\mathrm{ph}} .
    \label{eq:app_cross_dense}
\end{equation}

The corresponding ratios to the compact digital layers are
\begin{align}
    \frac{\Fph}{\Flr}
    &=
    \frac{M}{2\Nb},
    \label{eq:app_ph_lr_ratio}\\
    \frac{\Fph}{\Fttm}
    &=
    \frac{D}{2\Nb}.
    \label{eq:app_ph_tt_ratio}
\end{align}
At matched $P$, the crossover with the low-rank implementation therefore occurs
at
\begin{equation}
    \Nb = \frac{M}{2},
    \label{eq:app_cross_lr}
\end{equation}
while the crossover with the TTM implementation occurs at
\begin{equation}
    \Nb = \frac{D}{2}.
    \label{eq:app_cross_tt}
\end{equation}
For the binary tensorization $D=2$, the FLOP cost of constructing $BA$ once is
equal, at matched parameter count, to the cost of applying the TTM layer to one
dense input vector.  Any further reuse reduces the amortized photonic
electronic cost relative to the TTM contraction.

These relations are FLOP-count comparisons only.  They do not imply equal
wall-clock efficiency, because dense GEMMs, thin GEMMs, and TTM contractions
can attain very different fractions of the peak throughput of a CPU or GPU.

\subsection{Memory accounting}
\label{app:memory}

Let $s$ denote the number of bytes used to store one parameter
($s=4$ for FP32, $s=2$ for FP16/BF16).  The persistent parameter storage is
\begin{align}
    S_{\mathrm{dense}} &= sM^2,\\
    S_{\mathrm{LR}} &= sP_{\mathrm{LR}},\\
    S_{\mathrm{TTM}} &= sP_{\mathrm{TTM}},\\
    S_{\mathrm{ph}} &= sP_{\mathrm{ph}}.
\end{align}
The photonic representation therefore retains the same compact persistent
electronic state as an ordinary factorized representation with the same
parameter count.

Computing Eq.~\eqref{eq:app_ph_ba} also produces an $M\times M$ matrix.
If this result is materialized in memory, the product itself occupies
\begin{equation}
    S_{BA,\mathrm{out}} = sM^2 .
    \label{eq:app_ba_output_memory}
\end{equation}
Accordingly, the $BA$ construction has compact \emph{persistent} parameter
storage but an $O(M^2)$ temporary/output memory footprint during
reprogramming.  The precise peak allocation depends on the BLAS backend,
workspace requirements, and whether the output is constructed monolithically
or in tiles.  The numerical benchmarks therefore report both persistent model
storage and measured peak temporary allocation where the runtime provides such
information.

For low-rank and TTM execution, transient activation memory is likewise
implementation and batch-size dependent.  In particular, the TT
representation may have a small persistent parameter count while creating
larger intermediate tensors during contraction.  Peak runtime memory is
therefore measured separately from parameter storage rather than inferred from
$P$ alone.

\subsection{Complexity trade-offs}
\label{app:complexity_tradeoffs}

\begin{figure}[t]
    \centering
    \includegraphics[width=0.64\linewidth]
    {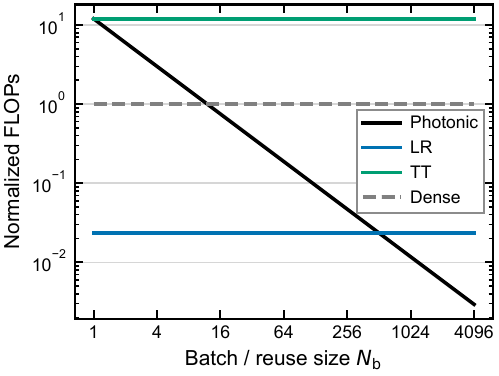}
    \caption{\textbf{Normalized electronic FLOP count versus batch/reuse
    size.}
    Leading-order arithmetic cost relative to dense $M\times M$
    multiplication for $M=1024$ and a representative compact parameter
    budget $P=2Mr$ with $r=12$.  Low-rank (LR) and tensor-train-matrix (TT)
    costs are incurred for every processed vector.  The photonic curve
    includes only the electronic construction of $BA$, performed once per
    reprogramming; subsequent optical matrix--vector multiplication is
    excluded.  The dense reference is normalized to unity.}
    \label{fig:app_flops_reuse}
\end{figure}

\begin{figure}[t]
    \centering
    \includegraphics[width=0.64\linewidth]
    {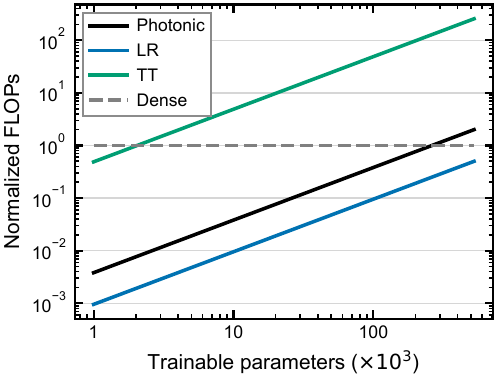}
    \caption{\textbf{Normalized electronic FLOP count versus trainable
    parameter count.}
    Leading-order electronic arithmetic cost relative to dense multiplication
    for $M=1024$, binary tensorization $D=2$, and
    $N_{\mathrm b}=128$.  The compact representations scale linearly with
    their trainable-parameter count $P$, but with different prefactors.
    The photonic cost corresponds exclusively to forming $BA$ once for a new
    configuration.}
    \label{fig:app_flops_parameters}
\end{figure}

The preceding expressions separate two distinct costs.  Digital structured
layers must evaluate their factorized representation for every input vector,
whereas the electronic cost associated with the photonic layer considered here
is incurred when a new optical configuration is constructed.  The latter cost
can therefore be amortized over all vectors processed before the next
reprogramming.  The following comparisons illustrate this distinction for
$M=1024$.  Unless stated otherwise, the compact representations are compared
at matched trainable-parameter count $P$, and the tensor-train baseline uses
binary tensorization ($D=2$).

Figure~\ref{fig:app_flops_reuse} shows the leading electronic FLOP count,
normalized to multiplication by a dense $M\times M$ matrix, as a function of
the number $N_{\mathrm b}$ of vectors processed with one configuration.  The
normalized costs of the low-rank and TTM layers are independent of
$N_{\mathrm b}$, since both their arithmetic cost and the dense reference scale
linearly with the number of input vectors.  In contrast, the cost of
constructing the photonic control matrix is paid once per reprogramming, so
its normalized contribution decreases as $1/N_{\mathrm b}$.

For the representative photonic parameterization $P=2Mr_{\mathrm{ph}}$, the ratio to dense electronic multiplication reduces to
    \begin{equation}
        \frac{F_{\mathrm{ph}}}{F_{\mathrm{dense}}}=
        \frac{r_{\mathrm{ph}}}{N_{\mathrm b}}.
    \end{equation}
Thus, for $r_{\mathrm{ph}}=12$, the arithmetic required to construct a new photonic configuration becomes smaller than the arithmetic required to apply the dense matrix once the same configuration is reused for more than approximately twelve input vectors.  This comparison concerns only the electronic construction of the control matrix; optical propagation and the physical programming of the interferometer are not included.

The dependence on the trainable-parameter budget is shown in
Figure~\ref{fig:app_flops_parameters}.  At fixed reuse $N_{\mathrm b}$, all three
compact representations have an electronic arithmetic cost that is linear in
$P$, but with markedly different coefficients:

\begin{equation}
    F_{\mathrm{LR}} = 2N_{\mathrm b}P,
    \qquad
    F_{\mathrm{TTM}}
    \simeq
    \frac{2N_{\mathrm b}M}{D}P,
    \qquad
    F_{\mathrm{ph}} = MP.
\end{equation}

The relative ordering is therefore governed not only by parameter count but
also by how the compact representation is evaluated.  In particular, the TTM
representation can require very little persistent parameter storage while
still carrying a substantial arithmetic cost when applied to an explicit dense
activation vector.  Conversely, the electronic construction cost of the
photonic layer is independent of the number of vectors subsequently processed
with the resulting optical configuration.

\begin{figure}[t]
    \centering
    \includegraphics[width=0.64\linewidth]
    {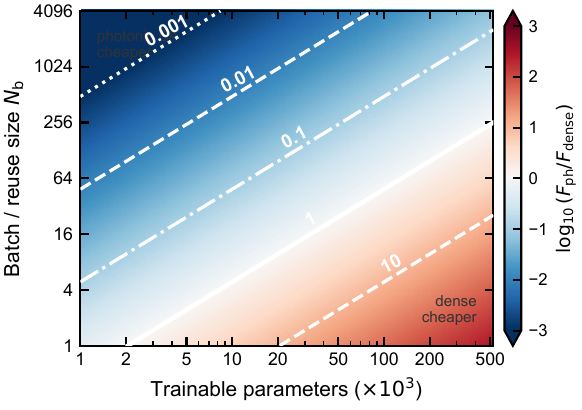}
    \caption{\textbf{Parameter--reuse map of the photonic electronic
    overhead.}
    Ratio $F_{\mathrm{ph}}/F_{\mathrm{dense}}$ for $M=1024$ as a function of
    trainable-parameter count $P$ and batch/reuse size $N_{\mathrm b}$.
    White contours indicate ratios
    $10^{-3}$, $10^{-2}$, $10^{-1}$, $1$, and $10$.
    Values below unity correspond to a $BA$ construction cost smaller than
    the FLOP count of dense electronic multiplication of the same
    $N_{\mathrm b}$ input vectors.  Optical execution and physical
    reprogramming overheads are not included.}
    \label{fig:app_parameter_reuse_map}
\end{figure}

The same dependence can be represented more directly in the
$(P,N_{\mathrm b})$ plane.  Figure~\ref{fig:app_parameter_reuse_map} shows

\begin{equation}
    R_{\mathrm{ph/dense}}
    =
    \frac{F_{\mathrm{ph}}}{F_{\mathrm{dense}}}
    =
    \frac{P}{2MN_{\mathrm b}}.
\end{equation}

The contour $R_{\mathrm{ph/dense}}=1$ separates the region in which forming
$BA$ requires fewer electronic FLOPs than dense multiplication of all
$N_{\mathrm b}$ vectors from the region in which it requires more.  The
additional contours quantify progressively stronger arithmetic amortization.
For example, $R_{\mathrm{ph/dense}}=10^{-2}$ means that the electronic
construction of the photonic configuration accounts for one hundredth of the
FLOPs required by the corresponding dense electronic multiplications.

The diagonal structure of the map makes the underlying scaling transparent:
increasing the trainable parameter budget increases the reprogramming cost,
whereas increasing reuse reduces its amortized contribution by the same
factor.  The relevant operating regime therefore depends on both model
compression and the number of vectors processed between parameter updates.

Analytical FLOP counts do not directly determine wall-clock execution time, because dense GEMMs, thin matrix products, and staged tensor contractions can use conventional processors with very different efficiency. We therefore benchmark the corresponding PyTorch implementations on a workstation equipped with an AMD Ryzen~9 3900X CPU, 64~GB RAM, and an NVIDIA GeForce RTX~3080 GPU, using FP32 arithmetic. For the dense, low-rank, and TTM layers, the measured time includes processing a batch of $N_{\mathrm b}$ dense activation vectors. For the photonic layer, it includes only a single construction of the latent control matrix $Z=BA$ and is therefore independent of $N_{\mathrm b}$. Phase transfer, device settling, calibration, optical propagation, modulation, and detection are excluded.

\begin{figure}[t]
    \centering
    \includegraphics[width=0.95\linewidth]
    {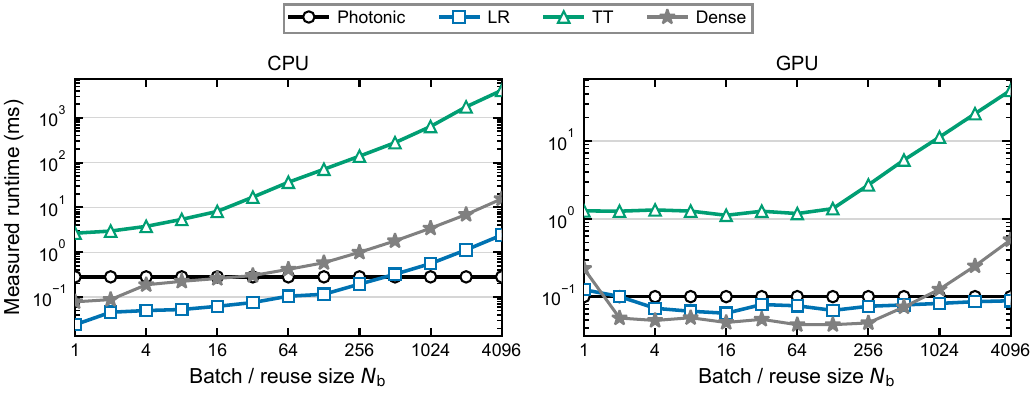}
    \caption{\textbf{Measured electronic runtime on CPU and GPU.}
    PyTorch runtime for dense, low-rank, and standard paired-core TTM
    multiplication as a function of batch/reuse size $N_{\mathrm b}$, compared
    with the time required to form $BA$ once for the photonic layer. Compact
    layers use approximately matched trainable-parameter budgets, with
    $M=1024$ and $r_{\mathrm{ph}}=r_{\mathrm{LR}}=12$. Measurements were
    performed on a workstation equipped with an AMD Ryzen 9 3900X CPU,
    64~GB RAM, and an NVIDIA GeForce RTX 3080 GPU, using FP32 arithmetic.
    GPU timings use CUDA events with explicit synchronization; host--device
    transfer is excluded. The photonic curve represents electronic
    control-matrix construction only and does not include interferometer
    programming or optical execution.}
    \label{fig:app_runtime}
\end{figure}

Figure~\ref{fig:app_runtime} shows that measured runtime can differ strongly from the nominal FLOP scaling, particularly on the GPU. For the dense layer, the $N_{\mathrm b}=1$ case behaves as a poorly utilized matrix--vector operation. Increasing $N_{\mathrm b}$ turns it into a more efficient GEMM, so runtime can initially decrease despite the larger operation count. Once the GPU is sufficiently utilized, at batch sizes of order a few hundred in the present benchmark, runtime begins to grow with $N_{\mathrm b}$.

The TTM layer exhibits a broader approximately constant-runtime regime at small and intermediate $N_{\mathrm b}$. Its staged contraction consists of several smaller matrix products and tensor-layout operations, so kernel-launch overhead and underutilization dominate initially. Increasing $N_{\mathrm b}$ improves the efficiency of these kernels before arithmetic and memory traffic become dominant, after which the runtime increases.

The photonic curve has a different interpretation: the $BA$ product is a one-time reprogramming cost. Reusing the same optical configuration therefore does not increase its electronic construction time, whereas the digital reference layers must process every additional activation vector. This illustrates why parameter count, FLOP count, and measured runtime should be treated as distinct quantities.

\begin{figure}[t]
    \centering
    \includegraphics[width=0.64\linewidth]
    {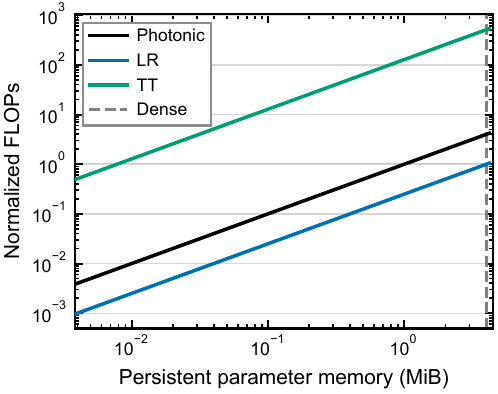}
    \caption{\textbf{Persistent-memory--compute trade-off.} Leading-order electronic FLOP count, normalized to dense multiplication, versus persistent trainable-parameter memory for $M=1024$, $N_{\mathrm b}=128$, $D=2$, and FP32 storage. The vertical dashed line denotes the parameter memory of the dense $M\times M$ matrix, while the horizontal unity level gives its normalized arithmetic cost.  Memory associated with temporary tensors, the materialized $BA$ product, and hardware control buffers is not included.}
    \label{fig:app_memory_compute}
\end{figure}

Finally, Figure~\ref{fig:app_memory_compute} relates arithmetic cost to persistent model storage.  The horizontal axis includes only trainable parameters.  For an $s$-byte numerical representation, the compact layers require $sP$ bytes, whereas a dense $M\times M$ matrix requires $sM^2$ bytes. The same reduction in trainable state also propagates to gradients and optimizer variables during training, although their precise memory footprint depends on the optimizer and numerical precision.

This persistent-memory comparison should be distinguished from peak working memory.  In particular, explicitly forming $BA$ produces an $M\times M$ matrix and therefore requires an $O(M^2)$ output buffer during reprogramming.  Similarly, tensor contractions may produce intermediate tensors considerably larger than the stored TTM cores.  Figure~\ref{fig:app_memory_compute} therefore characterizes the trade-off between persistent trainable state and arithmetic cost, rather than total peak system memory.

\section{Interferometer and optical interface details}

\subsection{Clements interferometer}\label{app:mesh_details}

As a particular unitary mesh, we consider the rectangular architecture of Clements \emph{et al.}~\cite{clements2016interferometers}, shown in Figure~\ref{fig:multipliers_types}b. The $M$-mode Clements interferometer comprises $M$ layers of MZIs acting on neighboring modes, $M(M-1)/2$ MZIs in total. Each MZI consists of two balanced (50:50) couplers and two phase shifters: the internal phase sets $\psi^{(l)}_{2j-1}$ the splitting ratio, and the external phase sets $\psi^{(l)}_{2j}$ the relative phase of the two modes. A column of $d$ single-mode phase shifters completes the circuit.

The transfer matrix of the mesh is
    \begin{equation}\label{eqn:clements_layers}
        U_\text{MZI}(\boldsymbol{\psi})=\Phi(\boldsymbol{\psi}^{(M+1)})\prod_{l=1}^{M}\,V^{(l)}_{\text{MZI}}(\boldsymbol{\psi}^{(l)}),
    \end{equation}
where $\Phi(\boldsymbol{\psi}^{(M+1)})=\mathrm{diag}\!\big(e^{i\psi^{(M+1)}_1},\ldots,e^{i\psi^{(M+1)}_d}\big)$ is the diagonal output phase layer and $V^{(l)}_{\text{MZI}}(\boldsymbol{\psi}^{(l)})$ is the $l$-th MZI layer. Each layer is a product of commuting nearest-neighbor blocks,
\[
    V^{(l)}_{\text{MZI}}=\prod_{j\in\Omega^{\text{MZI}}_l}T^{(l)}_{\text{MZI},j}\!\big(\psi^{(l)}_{2j-1},\psi^{(l)}_{2j}\big),
\]
where $T^{(l)}_{\text{MZI},j}$ equals the identity except for a single $2\times 2$ block acting on modes $j$ and $j+1$. In our convention this block reads
    \begin{equation}\label{eqn:mzi_block}
        B_{\text{MZI}}\!\big(\psi^{(l)}_{2j-1},\psi^{(l)}_{2j}\big)=\left(
        \begin{array}{cc}
            e^{i\psi^{(l)}_{2j}}\sin\!\left(\frac{\psi^{(l)}_{2j-1}}{2}\right) & \cos\!\left(\frac{\psi^{(l)}_{2j-1}}{2}\right) \\
            e^{i\psi^{(l)}_{2j}}\cos\!\left(\frac{\psi^{(l)}_{2j-1}}{2}\right) & -\sin\!\left(\frac{\psi^{(l)}_{2j-1}}{2}\right)
        \end{array}
        \right).
    \end{equation}
Together with the $M$ output phases $\boldsymbol{\psi}^{(M+1)}$, the mesh contains $M^2$ programmable phase shifts, matching the $M^2$ real degrees of freedom of the unitary group. A universal circuit therefore cannot be specified by fewer independent parameters. For notational convenience, we arrange the complete set of $M^2$ phase
shifts in a fixed order into an $M\times M$ array $\Psi$, defined by
    \begin{equation}
        \operatorname{vec}(\Psi)
        =
        \big(
            \boldsymbol{\psi}^{(1)},
            \boldsymbol{\psi}^{(2)},
            \ldots,
            \boldsymbol{\psi}^{(M+1)}
        \big),
        \label{eqn:phase_array}
    \end{equation}
where the right-hand side denotes concatenation of the phase vectors. We use $U_{\mathrm{MZI}}(\Psi)$ as shorthand for the transfer matrix of the mesh programmed with these phase values. With the real-valued encoding described in Appendix~\ref{app:optical_encoding}, the corresponding neural linear transformation is obtained from $U_{\mathrm{MZI}}(\Psi)$.

Universality therefore comes at full parameter cost: programming the mesh requires $M^2$ independent phase settings, the same number as the entries of a dense digital weight matrix. An MVM implemented in this way accelerates the multiplication itself, but the overheads that motivate parameter-efficient layers in the digital domain reappear in analog form. All $M^2$ phase values must be stored, updated at every optimization step, and delivered to the chip by per-phase drive electronics, and the calibration and reprogramming effort grows with the number of active elements~\cite{harris2018pnp,Song_64x64_mzi}. The construction described next decouples the number of trainable parameters from the number of physical phase shifters while preserving the expressivity of the universal mesh.

\subsection{Real-valued encoding and detection}\label{app:optical_encoding}

    \begin{figure}[t]
        \centering
        \includegraphics[width=0.95\linewidth]{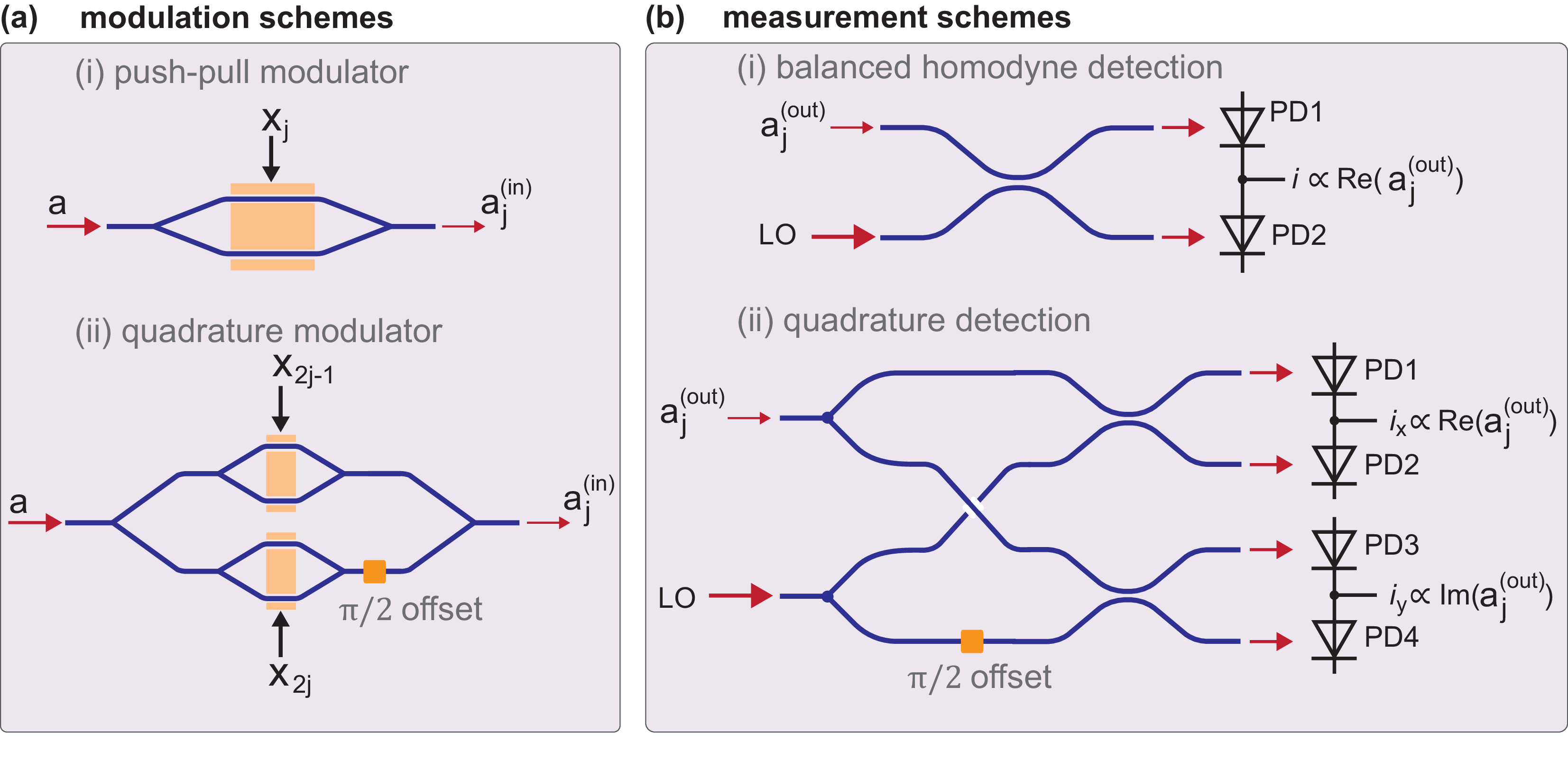}
        \caption{\textbf{Optical implementation of the two encodings}. (a)~Encoding (i): a push--pull amplitude modulator prepares the real input amplitude $a^{(\mathrm{in})}_j=x_j$; at the mesh output, balanced homodyne detection---mixing $a^{(\mathrm{out})}_j$ with a strong local oscillator (LO) on a balanced coupler and subtracting the photocurrents of the two photodiodes---yields $y_j\propto\mathrm{Re}\,a^{(\mathrm{out})}_j$. (b)~Encoding (ii): a quadrature modulator (two push--pull modulators in the arms of a Mach--Zehnder structure with a static $\pi/2$ offset) prepares $a^{(\mathrm{in})}_j=x_{2j-1}+\mathrm{i}\,x_{2j}$; each output field is split between two balanced homodyne receivers with LO phases offset by $\pi/2$, yielding $y_{2l-1}\propto\mathrm{Re}\,a^{(\mathrm{out})}_l$ and $y_{2l}\propto\mathrm{Im}\,a^{(\mathrm{out})}_l$.}
        \label{fig:encodings}
    \end{figure}

The interferometric mesh functions within the domain of complex-valued unitary matrices $U_{\mathrm{MZI}}$, facilitating the transformation of the vector of complex field amplitudes $\mathbf{a}^{(\mathrm{in})}\in\mathbb{C}^{M}$ introduced into its input ports, as expressed by
    \begin{equation}
        \mathbf{a}^{(\mathrm{out})}=U_{\mathrm{MZI}}(\Psi)\,\mathbf{a}^{(\mathrm{in})},
    \end{equation}
In contrast, the input and output signals of conventional neural-network layers are generally real-valued. Consequently, the optical multiplier necessitates an encoding scheme to map the real-valued input vector $\mathbf{x}$ onto the optical amplitudes $\mathbf{a}^{(\mathrm{in})}$, alongside a measurement scheme to retrieve the real-valued output vector $\mathbf{y}$ from $\mathbf{a}^{(\mathrm{out})}$. This requirement gives rise to two primary encoding methods for employing unitary interferometers as real-valued matrix multipliers. These encodings are depicted in Figure~\ref{fig:encodings}, which illustrates the optical modulation schemes employed to encode $\mathbf{x}$ into the input fields in panel (a) and the corresponding measurement schemes utilized to convert the output optical fields back into $\mathbf{y}$ in the electronic domain in panel (b).

(i) \textbf{One vector element per field amplitude}. In the first encoding scheme, each optical field amplitude carries a single real-valued vector element. The input fields are prepared with real amplitudes, $a^{(\mathrm{in})}_j=x_j$ ($j=1,\ldots,M$), while each output field is measured in a single quadrature, $y_l=\mathrm{Re}(a^{(\mathrm{out})}_l)$ ($l=1,\ldots,M$). The resulting effective real-valued matrix $W$ in \eqref{eqn:linear_map} is therefore 
    \begin{equation}\label{eqn:encoding_1b}
        W(\Psi)=\mathrm{Re}\!\left(U_{\mathrm{MZI}}(\Psi)\right).
    \end{equation}
Notice that in \eqref{eqn:encoding_1} the $M\times{}M$ matrix $W=\mathrm{Re}\!\left(U_{\mathrm{MZI}}(\Psi)\right)$ is not constrained to be unitary. Physically, to attain the encoding, the input amplitudes can be set using push--pull amplitude modulators (scheme (i) in Figure~\ref{fig:encodings}a). The output fields are measured by balanced homodyne detection (scheme (i) in Figure~\ref{fig:encodings}b), for which each output field is combined with a strong local oscillator on a balanced coupler, and the difference between the two resulting photocurrents is proportional to the field quadrature that is in phase with the local oscillator.

(ii) \textbf{Two vector elements per field amplitude}. In the second encoding scheme, each complex field amplitude represents two real-valued vector elements, one in each quadrature. The input amplitudes are encoded as $a^{(\mathrm{in})}_j=x_{2j-1}+\mathrm{i}x_{2j}$, while the output components are recovered according to $y_{2l-1}=\mathrm{Re}\left(a^{(\mathrm{out})}_l\right)$ and $y_{2l}=\mathrm{Im}\left(a^{(\mathrm{out})}_l\right)$ ($j,l=1,\ldots,M$). An interferometric mesh of $M$-mode can, therefore, implement a transformation on real vectors of dimension $2M$.  The two input quadratures are prepared using a quadrature modulator, in which two push--pull amplitude modulators operate in parallel in the two arms of a Mach--Zehnder modulators with a fixed $\pi/2$ phase offset between the arms (scheme (ii) in Figure~\ref{fig:encodings}a). At the output, each field is split equally between two balanced homodyne receivers whose local oscillators differ in phase by $\pi/2$, thereby allowing the two quadratures to be measured simultaneously (scheme (ii) in Figure~\ref{fig:encodings}b).

The corresponding effective $2M\times2M$ real representation of $U_{\mathrm{MZI}}(\Psi)$ is an orthogonal matrix, consisting of $2\times2$ blocks representing the entries of $U$:
    \begin{equation}\label{eqn:realification}
        W(\Psi)=\mathrm{Re}\,(U(\Psi))\otimes I_{2}
               +\mathrm{Im}\,(U(\Psi))\otimes J,
        \qquad
        J=\begin{pmatrix}0&-1\\1&0\end{pmatrix},
    \end{equation}
so that unitarity of $U(\Psi)$ implies $W^{T}W=I_{2M}$.

The two schemes trade transceiver simplicity against the structure of the effective map. The first requires a single modulator and a single homodyne receiver per vector element. The price is that $\mathrm{Re}\,(U)$ is not norm preserving---the maps reachable in this scheme are exactly the real matrices of spectral norm at most unity. The second doubles the transceiver hardware, but its orthogonal map preserves the Euclidean norm of the data through the layer, which supports stable signal propagation in deep networks.

\section{Details of baseline parameter-efficient layers}
\label{app:baseline_layers}

This Appendix gives the explicit constructions of the digital
parameter-efficient layers summarized in
Sec.~\ref{sec:baseline_layers}. The notation follows that section, with
square transformations of dimension $M$ unless stated otherwise.

\subsection{Low-rank layer}

The low-rank baseline factorizes
\begin{equation}
    W^{(\mathrm{LR})}=BA,
    \qquad
    B\in\mathbb{R}^{M\times r},
    \qquad
    A\in\mathbb{R}^{r\times M},
\end{equation}
and therefore contains
\begin{equation}
    P_{\mathrm{LR}}=2Mr
\end{equation}
trainable parameters. In end-to-end training this factorization replaces the
dense weight matrix directly. In parameter-efficient fine-tuning the same
construction is used as an additive update to a frozen pretrained matrix,
corresponding to LoRA~\cite{hu2021lora}.

\subsection{Tensor-network representation}

For the tensor-network baselines, we factor the layer dimension as
\begin{equation}
    M=D^Q
\end{equation}
and reshape
$W\in\mathbb{R}^{M\times M}$ into the $2Q$-way tensor
\begin{equation}
    \mathcal{W}_{i_1,\ldots,i_Q,j_1,\ldots,j_Q},
    \qquad
    i_q,j_q\in\{1,\ldots,D\}.
\end{equation}
The corresponding linear map is
\begin{equation}
    Y_{i_1\ldots i_Q}
    =
    \mathcal{W}_{i_1,\ldots,i_Q,j_1,\ldots,j_Q}
    X_{j_1\ldots j_Q},
\end{equation}
with summation over repeated indices. Compression is obtained by expressing
$\mathcal{W}$ as a contraction of lower-order tensors.

\subsection{Tensor-train layer}

We use the TT-matrix representation of
Ref.~\cite{novikov2015tensorizing}. The weight tensor is expressed through
four-way cores
\begin{equation}
    \mathcal{G}^{(q)}
    \in
    \mathbb{R}^{r_{q-1}\times D\times D\times r_q},
\end{equation}
with TT ranks satisfying $r_0=r_Q=1$:
\begin{equation}
    \mathcal{W}_{i_1,\ldots,i_Q,j_1,\ldots,j_Q}
    =
    \sum_{\alpha_0,\ldots,\alpha_Q}
    \prod_{q=1}^{Q}
    \mathcal{G}^{(q)}_
    {\alpha_{q-1},i_q,j_q,\alpha_q}.
    \label{eq:app_tt_contraction}
\end{equation}
The number of trainable parameters is
\begin{equation}
    P_{\mathrm{TT}}
    =
    D^2
    \sum_{q=1}^{Q}
    r_{q-1}r_q .
    \label{eq:app_tt_params_baselines}
\end{equation}
The parameter budget is varied by changing the TT ranks. We use the shorter
term ``TT layer'' in the main text, although the paired input--output-core
representation in Eq.~\eqref{eq:app_tt_contraction} is more precisely a
TT-matrix representation.

\subsection{Brickwork tensor-network layer}

The brickwork layer uses a depth-$L$ tensor network composed of four-way cores
acting on pairs of neighboring tensor modes. Successive layers shift the
pairing pattern, producing the alternating local connectivity illustrated in
Fig.~\ref{fig:digital_layers}(c).

The full weight tensor is written as
\begin{equation}
    \mathcal{W}
    =
    T^{(1)}T^{(2)}\cdots T^{(L)},
    \label{eq:app_bw_factorization}
\end{equation}
where contraction over the intermediate tensor indices is implicit. Each
layer tensor $T^{(l)}$ factorizes into
\begin{equation}
    G^{(l,k)}
    \in
    \mathbb{R}^{D\times D\times D\times D}
\end{equation}
acting on pairs of neighboring modes. For even $Q$, odd layers pair
$(1,2),(3,4),\ldots$, while even layers use the shifted pairing
$(2,3),(4,5),\ldots$ together with the boundary pair $(Q,1)$.

Since each of the $L$ layers contains $Q/2$ four-way cores, the number of
trainable parameters is
\begin{equation}
    P_{\mathrm{BW}}
    =
    \frac{Q}{2}LD^4
    =
    \frac{1}{2}LD^4\log_D M .
    \label{eq:app_bw_params}
\end{equation}
The parameter budget in the numerical experiments is varied through the
brickwork depth $L$.

\subsection{Orthogonal butterfly fine-tuning}

For multiplicative parameter-efficient fine-tuning we use
BOFT~\cite{BOFT}, implemented with the BOFT adapter from the Hugging
Face PEFT library. BOFT constructs an orthogonal transformation from sparse
block-orthogonal factors connected by fixed butterfly permutations.

For an input feature dimension $d$, each factor contains $d/b$ orthogonal
blocks of size $b\times b$. If $m$ butterfly factors are used, the effective
number of independent orthogonal degrees of freedom is
\begin{equation}
    P_{\mathrm{BOFT}}^{(\mathrm{dof})}
    =
    \frac{1}{2}(b-1)dm ,
    \label{eq:app_boft_params}
\end{equation}
where $b$ is the block size. Thus, for fixed $d$ and $m$, increasing $b$
increases the capacity of the adapter.

Schematically, the transformation can be written as
\begin{equation}
    R
    =
    \prod_{j=1}^{m}
    P_j
    \operatorname{diag}
    \left(
        Q_{j,1},\ldots,Q_{j,d/b}
    \right)
    P_j^{\mathsf T},
    \qquad
    Q_{j,k}^{\mathsf T}Q_{j,k}=I ,
    \label{eq:app_boft}
\end{equation}
where $P_j$ denotes a fixed butterfly permutation and $Q_{j,k}$ is an
orthogonal block generated from trainable variables using the Cayley
parameterization.

The BOFT adapter acts multiplicatively on a frozen pretrained layer. With the
matrix convention used in this work,
\begin{equation}
    W_{\mathrm{adapt}}
    =
    W_0R .
\end{equation}
The input feature dimension is denoted by $d$ rather than $M$ because the
pretrained layer need not be square. In the experiments, we vary
\texttt{boft\_block\_size} to change the BOFT parameter budget while keeping
the remaining configuration fixed. The parameter counts reported in the
figures are the actual trainable scalar counts of the instantiated PEFT
adapters; Eq.~\eqref{eq:app_boft_params} instead gives the independent
orthogonal degrees of freedom.

\section{Sequence-memorization benchmark with Qwen3}
\label{app:qwen_memorization}

This Appendix provides details of the sequence-memorization benchmark used in
Sec.~\ref{sec:results_memorization}. The purpose of the benchmark is to test
the capacity of parameter-efficient updates to store information that is not
contained in the pretrained model. Following Ref.~\cite{lora_memory}, we use
exact memorization of synthetic token sequences as a controlled measure of
parametric memory. Synthetic data are used in order to minimize the influence
of linguistic structure, semantic associations, and knowledge already present
in the pretrained model.

\subsection{Memorization task}
\label{app:qwen_task}

We use the pretrained Qwen3-0.6B model~\cite{qwen}. A language model represents
text as a sequence of discrete tokens drawn from a finite vocabulary. For the
present benchmark, rather than using natural-language text, we generate
$N=16$ random token sequences,
\begin{equation}
    \mathbf{s}^{(i)}
    =
    \left(
    s^{(i)}_1,\ldots,s^{(i)}_{\ell}
    \right),
    \qquad
    i=1,\ldots,N,
\end{equation}
each of length $\ell=200$.

During fine-tuning, the pretrained model parameters are kept fixed and only the
parameter-efficient adapters are optimized to store the generated sequences.
During evaluation, the model is supplied with the beginning of a stored
sequence and is required to reproduce the remaining tokens autoregressively.
Generation uses greedy decoding: at each position the token assigned the
highest probability by the model is selected. The resulting generated sequence
is denoted
\begin{equation}
    \hat{\mathbf{s}}^{(i)}
    =
    \left(
    \hat{s}^{(i)}_1,\ldots,\hat{s}^{(i)}_{\ell}
    \right).
\end{equation}

This task differs from conventional downstream language-model benchmarks in
that successful performance does not require semantic understanding or
generalization to unseen examples. Instead, it probes how efficiently the
trainable parameters can encode and recover a prescribed finite data set.

\subsection{Adapter placement and parameterization}
\label{app:qwen_adapters}

Although the input and output of the task are discrete tokens, the transformer operates internally on continuous activation vectors. The same linear transformations are applied independently to the activations associated with each token position. Consequently, the sequence length controls the number of times a linear layer is evaluated but does not change the dimension of the corresponding matrix--vector multiplication.

In the benchmark, parameter-efficient adapters are introduced into the \texttt{down\_proj} modules of transformer blocks $15$ and $19$, while all pretrained Qwen3-0.6B parameters remain frozen (see Figure~\ref{fig:architectures}c). The native down-projection matrix has dimensions $1024\times3072$. The parameter-efficient constructions compared in the benchmark implement $1024\times1024$ trainable transformations $\Delta W$ associated with these modules.

For the proposed photonic adapter, the $1024$-dimensional real activation is represented using the norm-preserving encoding (ii) of Sec.~\ref{app:optical_encoding}. Two real components are encoded in each complex optical amplitude, so that the corresponding unitary interferometer contains $M=512$ optical modes. With the factorized control map used throughout this work, one photonic adapter therefore contains
    \begin{equation}
        P_{\mathrm{ph}}^{(1)}=2rM=1024r
    \end{equation}
trainable parameters. Since adapters are inserted into two transformer blocks, the total photonic parameter count is
    \begin{equation}
        P_{\mathrm{ph}}=2048r.
    \end{equation}

The LoRA and BOFT baselines are applied to the same two locations. For the $1024\times1024$ transformation considered here, a rank-$r$ LoRA adapter contains $2r\times1024$ trainable parameters per adapted layer. The parameter counts reported in Fig.~\ref{fig:memo} are the totals over both adapted transformer blocks.

For reference, full fine-tuning updates the two native $1024\times3072$ down-projection matrices, corresponding to approximately $6.3\times10^{6}$ trainable parameters.

\subsection{Memorization metrics}
\label{app:qwen_metrics}

We quantify memorization using greedy recall accuracy and the first-fail position.

The \emph{greedy recall accuracy} is the fraction of target tokens reproduced correctly under greedy decoding,
\begin{equation}
    A_{\mathrm{rec}}
    =
    \frac{1}{N\ell}
    \sum_{i=1}^{N}
    \sum_{k=1}^{\ell}
    \delta_{s^{(i)}_k,\hat{s}^{(i)}_k},
    \label{eq:app_recall_accuracy}
\end{equation}
where $\delta$ is the Kronecker delta. Thus, $A_{\mathrm{rec}}=1$ corresponds to perfect reproduction of every token in the data set.

Token-averaged accuracy does not distinguish between isolated errors and an early error followed by a long incorrect continuation. We therefore also measure the position of the first decoding error. For sequence $i$,
    \begin{equation}
        f^{(i)}=
        \min
        \left\{
            k:
            \hat{s}^{(i)}_k\neq s^{(i)}_k
        \right\},
        \label{eq:app_first_fail}
    \end{equation}
with $f^{(i)}=\ell+1$ when the complete sequence is reproduced without error. The reported mean first-fail position is
    \begin{equation}
        \overline{f}
        =
        \frac{1}{N}
        \sum_{i=1}^{N}
        f^{(i)}.
        \label{eq:app_mean_first_fail}
    \end{equation}
Larger $\overline{f}$ therefore indicates that a longer prefix is reproduced exactly before the first error.

For completeness, we also define the \emph{greedy exact-match accuracy} as the fraction of sequences reproduced correctly in their entirety,
    \begin{equation}
        A_{\mathrm{exact}}
        =
        \frac{1}{N}
        \sum_{i=1}^{N}
        \prod_{k=1}^{\ell}
        \delta_{s^{(i)}_k,\hat{s}^{(i)}_k}.
        \label{eq:app_exact_match}
    \end{equation}
The first two metrics are used in Fig.~\ref{fig:memo}; the first-fail position provides a particularly sensitive measure of the onset of exact sequence recall.

\newpage

\bibliographystyle{unsrtnat}
\bibliography{refs}

\end{document}